\documentclass[aps,pra,twocolumn,showpacs,superscriptaddress,groupedaddress]{revtex4}
\usepackage[T1]{fontenc}
\usepackage[latin9]{inputenc}

\usepackage{mathrsfs}
\usepackage{amsmath}
\usepackage{graphicx}

\makeatletter
\usepackage{xcolor}

\@ifundefined{showcaptionsetup}{}{%
 \PassOptionsToPackage{caption=false}{subfig}}
\makeatother

\usepackage{babel}
\usepackage[colorlinks,citecolor=blue,linkcolor=blue]{hyperref}

\begin{document}
\title{Efficiency statistics of a quantum Otto cycle}
\author{Zhaoyu Fei}
\address{Graduate School of China Academy of Engineering Physics, No. 10 Xibeiwang
East Road, Haidian District, Beijing, 100193, China}
\author{Jin-Fu Chen}
\address{Graduate School of China Academy of Engineering Physics, No. 10 Xibeiwang
East Road, Haidian District, Beijing, 100193, China}
\address{Beijing Computational Science Research Center, Beijing 100193, China}
\address{School of Physics, Peking University, Beijing, 100871, China}
\author{Yu-Han Ma}
\email{yhma@gscaep.ac.cn}

\address{Graduate School of China Academy of Engineering Physics, No. 10 Xibeiwang
East Road, Haidian District, Beijing, 100193, China}
 \begin{abstract}
   The stochastic efficiency [G. Verley et al., Nat. Commun. \textbf{5}, 4721 (2014)] was introduced to evaluate the performance of energy-conversion machines in micro-scale. However, such an efficiency generally diverges when no heat is absorbed while work is produced in a thermodynamic cycle. As a result, any statistical moments of the efficiency do not exist. In this study, we come up with a different version of the definition for the stochastic efficiency which is always finite. Its mean value is equal to the conventional efficiency, and higher moments characterize the fluctuations of the cycle. In addition, the fluctuation theorems are re-expressed via the efficiency. For working substance satisfying the equipartition theorem, we clarify that the thermodynamic uncertainty relation for efficiency is  valid in an Otto engine. To demonstrate our general discussions, the efficiency statistics of a quantum harmonic-oscillator Otto engine is systematically investigated. The probability that the stochastic efficiency surpasses the Carnot efficiency is explicitly obtained.  This work may shed new insight for optimizing micro-machines with fluctuations.
 \end{abstract}

\maketitle
\section{Introduction}

For a heat engine operating between a hot and
cold reservoir, the conventional efficiency is defined by the ratio of the output work and the heat absorbed from the hot reservoir, which characterizes the performance of the engine.
As the size of the engine decreases, the thermal fluctuations~\cite{seifert2012, seikimoto2010} and quantum fluctuations~\cite{esposito2009,campisi2011} become  more significant. From the point view of stochastic thermodynamics,
the work, heat, and entropy of microscopic systems are all stochastic
quantities.  Hence, at a microscopic level, it is natural to expect that the efficiency, introduced to evaluate
the ability of energy conversion for various thermal machines, is
also a stochastic quantity. %In this sense, the efficiency fluctuation,
%which is ignored in macroscopic systems, should be considered to describe
%the performance of microscopic thermal machines.

Recently, the stochastic efficiency, defined as the ratio of the stochastic output work and the stochastic heat absorbed from the hot reservoir in a cycle, has been widely studied for classical heat engines~\cite{verley2013, verley2014, mani2019, martinez2016, polett2015}.
This  stochastic efficiency is also applied to a quantum Otto cycle in Refs.~\cite{denzler2020,denzler20212}. However, such a definition of the stochastic efficiency seems weird
for the following three reasons: (1) The mean value of the stochastic
efficiency is not equal to the conventional efficiency in general. In contrary,
the mean values of stochastic work, heat, and entropy are equal to
their counterparts in the conventional thermodynamics; (2) The efficiency
approaches infinity with a non-zero probability. Such a result is due to the possibility that no heat is absorbed from the hot reservoir while work is produced in one realization of the cycle~\cite{denzler20212};
(3) Due to the divergent efficiency distribution, any moments
of the efficiency are ill-defined~\cite{polett2015}. Thus, one fails to evaluate the performance of the heat engine by the moments of this version of stochastic efficiency.

To avoid such weirdness, meanwhile to evaluate the fluctuations in a practical heat engine, we
come up with a different version of the stochastic efficiency. %We also require that
%mean value of the stochastic efficiency is equal to the thermal efficiency in consistent with other thermodynamic quantities in stochastic thermodynamics.
Then, the fluctuation theorems~\cite{seifert2012, seikimoto2010,esposito2009,campisi2011, jarzynski2011, crooks1999} is re-expressed via the stochastic efficiency. Moreover, the thermodynamic uncertainty relation (TUR)~\cite{barato2015,pietz2016} for efficiency is investigated. For working substance satisfying the equipartition theorem, we obtain the TUR for a quantum Otto cycle in the quasistatic limit (a general proof) and in a finite-time Otto cycle (numerical simulations).
As a specific example, we apply our version of the stochastic efficiency to study a quantum harmonic-oscillator Otto cycle. We find that both the probability that the stochastic efficiency surpasses the Carnot efficiency and the probability that the stochastic efficiency is negative increase as the temperatures of the reservoirs decrease.
%For a frequency conversion harmonic oscillator
%being the working substance of a quantum Otto cycle,  the statistics of the stochastic efficiency is investigated.  The stochastic efficiency exceeds the upper bound of the thermal efficiency (the Carnot efficiency) with a nonzero probability. Meanwhile, the probability of the engine to be useless is larger.

This paper is arranged as follows. In Sec.~\ref{s2}, we introduce the quantum Otto cycle and the joint distribution of input work and absorbed heat from the hot reservoir. In Sec.~\ref{s3}, a different version of stochastic efficiency is given.
The fluctuation theorems and the TUR are also re-expressed via the stochastic efficiency. In Sec.~\ref{s4}, we demonstrate our general discussions in a quantum  Otto cycle with the harmonic oscillator being the working substance. And we systematically investigate the statistics of the stochastic efficiency.
Section~\ref{s5} is the summary and discussion.

\section{the joint distribution of work and heat in a quantum Otto cycle}
\label{s2}

\begin{figure}
\includegraphics[width=8.5cm]{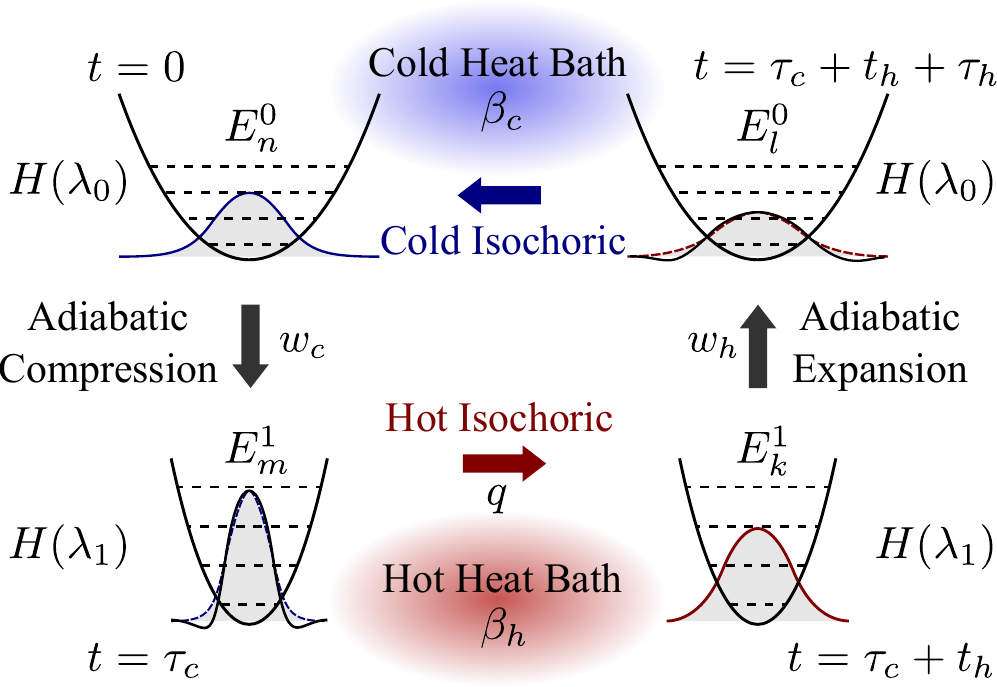}

\caption{\label{fig:Cycle} The schematic of a finite-time quantum Otto cycle. %The four processes in the cycle
%are the adiabatic compression process, the hot isochoric process,
%the adiabatic expansion process and the cold isochoric process. %We
%only require $w_{c},w_{e}$ and $q$ to define the fluctuating efficiency.
\label{fig:illustration}}

\end{figure}

As illustrated in Fig.~\ref{fig:Cycle}, we consider a quantum Otto cycle which involves four strokes:
two adiabatic processes and two isochoric processes~\cite{ht2007, rezek2006}.
In the adiabatic compression  (expansion) process, the Hamiltonian
of the working substance is changed from $H(\lambda_0)$
to $H(\lambda_1)$ (from $H(\lambda_1)$ to $H(\lambda_0)$) during
time $\tau_c$ ($\tau_h$) through a time-dependent parameter $\lambda$.
In the two isochoric processes, during time $t_h$ ($t_c$), the working
substance contacts a hot (cold) reservoir at the inverse
temperature $\beta_h$ ($\beta_c$) with fixed $\lambda$. For simplicity, we assume that
a complete thermalization is achieved in the two isochoric processes.
Namely, the working substance is thermal equilibrium with the corresponding
reservoir at the end of each isochoric process.

The stochastic work and heat in a quantum Otto cycle are defined under
the two-point measurement scheme~\cite{denzler2020}. At time $t=0,\tau_c,\tau_c+t_h,\tau_c+t_h+\tau_h$
($t=0$ is the initial time of the adiabatic compression process), we apply
the projective measurements of energy on the working substance according
to the corresponding instantaneous Hamiltonian. Then, the stochastic
work $w_c$ ($w_e$) in the adiabatic compression (expansion) process,
and the stochastic absorbed heat $q$ in the hot isochoric process
are defined as
\begin{align}
w_{c} & =E_{m}^{1}-E_{n}^{0}\nonumber \\
q & =E_{k}^{1}-E_{m}^{1}\label{eq:1}\\
w_{e} & =E_{l}^{0}-E_{k}^{1},\nonumber
\end{align}
 where $E_n^0,E_m^1,E_k^1,E_l^0$ are the measured energy of the four
projective measurements corresponding to the times $t=0,\tau_c,\tau_c+t_h,\tau_c+t_h+\tau_h$ respectively
($n,m,k,l$ denote the corresponding quantum numbers). Thus, the joint
probability distribution $P(w,q)$ of the total stochastic input work
$w=w_c+w_e$, and the stochastic absorbed heat from the hot reservoir $q$ is given by
\begin{align}
\ensuremath{P(w,q)} =&\sum_{n,m,k,l}\delta(w-E_{m}^{1}+E_{n}^{0}-E_{l}^{0}+E_{k}^{1})\delta(q-E_{k}^{1}+E_{m}^{1})\nonumber \\
 &\times \left|_{1}\left\langle m\left|U_{c}\right|n\right\rangle _{0}\right|^{2}\left|_{0}\left\langle l\left|U_{h}\right|k\right\rangle _{1}\right|^{2}\frac{e^{-\beta_{c}E_{n}^{0}-\beta_{h}E_{k}^{1}}}{Z^{0}(\beta_c)Z^{1}(\beta_h)},\label{eq:2}
\end{align}
where $U_c,U_h$ are the unitary evolution operators corresponding to
the compression and expansion processes, $|j\rangle_{0(1)}$ is the eigenstate of the Hamiltonian $H(\lambda_0)$ ($H(\lambda_1)$) and $Z^{0}(\beta_c)=\mathrm{Tr}[e^{-\beta_c H(\lambda_0)}]$,
$Z^{1}(\beta_h)=\mathrm{Tr}[e^{-\beta_h H(\lambda_1)}]$ are the partition
functions corresponding to the equilibrium states at $t=0$ and $t=\tau_c+t_h$ respectively.

\section{Stochastic efficiency}
\label{s3}

\subsection{Definition}

We define the stochastic efficiency of a (classical) quantum heat engine as
\begin{equation}
\eta=-\frac{w}{\left\langle q\right\rangle },\label{eq:3}
\end{equation}
where $\left\langle \cdot\right\rangle $ denotes the mean value over
numerous measurements, i.e.,
\begin{equation}
\left\langle q\right\rangle =\int dwdqP(w,q)q.\label{eq:4}
\end{equation}
For heat engines, the denominator
is always non-zero ($\left\langle q\right\rangle >0$), so the stochastic efficiency $\eta$ in Eq. (\ref{eq:3})
is finite. Moreover, it follows from Eq. (\ref{eq:4})
that $\left\langle \eta\right\rangle =-\left\langle w\right\rangle /\left\langle q\right\rangle $,
which is just the conventional efficiency.
%This definition of the stochastic efficiency is extended to classical systems straightforwardly.

From the joint distribution
$P(w,q)$, the distribution of the stochastic efficiency $P(\eta)$
is obtained by
\begin{equation}
P(\eta)=\int dwdqP(w,q)\delta(\eta+w/\left\langle q\right\rangle ).\label{eq:5}
\end{equation}
The fluctuation of the stochastic efficiency is determined by the output work, which characterizes the reliability of  the heat engine.

For practical calculation of the joint distribution of work and
 heat, we show the characteristic function of $P(w,q)$ in
the following
\begin{equation}
\chi(u,v)\equiv\left\langle e^{iuw+ivq}\right\rangle =\chi_{c}(u,v)\chi_{h}(u,v),\label{eq:12}
\end{equation}
where
\begin{equation}
\chi_{c}(u,v)=\frac{\mathrm{{Tr}}[U_{c}^{\dagger}e^{(iu-iv)H(\lambda_{1})}U_{c}e^{(-iu-\beta_{c})H(\lambda_{0})}]}{Z^{0}(\beta_c)},\label{eq:13-1}
\end{equation}
\begin{equation}
\chi_{h}(u,v)=\frac{\mathrm{{Tr}}[U_{e}^{\dagger}e^{iuH(\lambda_{0})}U_{e}e^{(-iu+iv-\beta_{h})H(\lambda_{1})}]}{Z^{1}(\beta_h)}.\label{eq:14-1}
\end{equation}
The cumulant moments of work and heat are obtained from $\chi(u,v)$, such as the average input work
\begin{equation}
\label{aw}
\left\langle w\right\rangle =\left.-i\frac{\partial\ln\chi(u,v)}{\partial u}\right|_{u=v=0},
\end{equation}
the average heat absorbed form the hot reservoir
\begin{equation}
\label{aq}
\left\langle q\right\rangle =\left.-i\frac{\partial\ln\chi(u,v)}{\partial v}\right|_{u=v=0},
\end{equation}
and the variance of  input work
\begin{equation}
\label{dw}
\Delta w^2 =\left.-\frac{\partial^2\ln\chi(u,v)}{\partial u^2}\right|_{u=v=0},
\end{equation}
where $\Delta(\cdot)=\sqrt{\left\langle \cdot^{2}\right\rangle -\left\langle \cdot\right\rangle ^{2}}$ denotes the standard deviation.

\subsection{Fluctuation theorems}

Fluctuation theorems indicate the equality relation in a general nonequilibrium
process. According to Ref.~\cite{sinit2011},  the fluctuation theorems are reexpressed via the efficiency for the Hamiltonian of the working substance involving time-reversal symmetry:

\begin{equation}
\left\langle e^{-\delta s(\eta,q)}\right\rangle =1,\label{eq:8}
\end{equation}
\begin{equation}
P(-\eta\left\langle q\right\rangle ,q)=P_{R}(\eta\left\langle q\right\rangle ,-q)e^{\delta s(\eta,q)},\label{eq:9}
\end{equation}
where $\delta s(\eta,q)=\beta_{c}(\eta_{C}q-\eta\left\langle q\right\rangle )$
is the total stochastic entropy production expressed in terms of $\eta$
and $q$, and $\eta_{C}=1-\beta_{h}/\beta_{c}$ is the Carnot efficiency.
The subscript $R$ denotes the reverse process of the cycle (the clockwise
direction in Fig.~\ref{fig:illustration}). Then, using the Jensen's inequality $e^{\left\langle x\right\rangle }\leq\left\langle e^{x}\right\rangle $,
we have $\left\langle \eta\right\rangle \leq\eta_{C}$ for heat engines
($\left\langle q\right\rangle >0$), which is the second law of thermodynamics.
It is worth mentioning that this inequality is not sharp. In fact,
for quantum systems without energy-level crossing when changing the
parameter $\lambda$, we obtain a sharper inequality $\left\langle \eta\right\rangle \leq\eta_{O}$ as a result of the minimum work principle~\cite{alla2005},
where $\eta_{O}$ is the Otto efficiency, i.e., the efficiency of an Otto cycle in the quasistatic limit (see Appendix A).

\subsection{Thermodynamic uncertainty relation}
\label{s3b}

Since the fluctuation theorems always imply the genralized TUR~\cite{timp2019}, it follows from Eq. (\ref{eq:9}) that
\begin{equation}
\frac{\mathrm{\Delta}\eta^{2}}{\left\langle \eta\right\rangle ^{2}}\geq f(\left\langle \delta s\right\rangle ),\label{eq:11}
\end{equation}
where $f(x)=\mathrm{csch}^2 [g(x/2)]$, and $g(x)$
is the inverse function of $x\mathrm{tanh}x$. Equation~(\ref{eq:11})
expresses a trade-off between the relative fluctuation of the efficiency
and the dissipation quantified through the entropy production in a
cycle. When $\left\langle \delta s\right\rangle\to 0$, $f(\left\langle \delta s\right\rangle )\approx 2/\left\langle \delta s\right\rangle$, which reproduces the TUR~\cite{barato2015,pietz2016} for the efficiency.

For the spectra of the working substance with scale property, i.e., $E_n^1=E_n^0/\epsilon$~\cite{denzler2020} ($\epsilon$ is $n$-independent), the general expression of the joint characteristic function (Eq.~(\ref{eq:12})) is obtained in the quasistatic limit (see Appendix B). From Eqs.~(\ref{aw}),(\ref{dw}), we obtain%the average input work $\left\langle w\right\rangle$ and the variance of input work $\Delta w^2$ become
%Then, the average input work $\left\langle w\right\rangle$, the average heat absorbed from the hot reservoir $\left\langle q\right\rangle$, the average entropy production of the cycle $\left\langle \delta s\right\rangle$, and the variance of input work $\Delta w^2$ become
%\begin{gather}
 %\begin{split}
%\label{e12}
%\left\langle w\right\rangle &=\left.-i\frac{\partial\ln\chi(u,v)}{\partial u}\right|_{u=v=0}=\eta_OT_c\left(\frac{\sigma_c}{1-\eta_O}-\frac{\sigma_h}{1-\eta_C}\right),\\
%\left\langle q\right\rangle &=\left.-i\frac{\partial\ln\chi(u,v)}{\partial v}\right|_{u=v=0}=-T_c\left(\frac{\sigma_c}{1-\eta_O}-\frac{\sigma_h}{1-\eta_C}\right),\\
%\left\langle \delta s\right\rangle &=\beta_c\left\langle q\right\rangle(\eta_C-\eta_O)=\frac{\eta_C-%\eta_O}{k_B}\left(\frac{\sigma_c}{1-\eta_O}-\frac{\sigma_h}{1-\eta_C}\right),\\
%\Delta w^2&=\left.-\frac{\partial^2\ln\chi(u,v)}{\partial u^2}\right|_{u=v=0}\\
%&=\eta_O^2k_B T_c^2 \left[\frac{C_c}{(1-\eta_O)^2}+\frac{C_h}{(1-\eta_C)^2}\right],
 %\end{split}
%\end{gather}
\begin{gather}
 \begin{split}
\label{e12}
\left\langle w\right\rangle &=\eta_OT_c\left(\frac{\sigma_c}{1-\eta_O}-\frac{\sigma_h}{1-\eta_C}\right),\\
\Delta w^2 &=\eta_O^2k_B T_c^2 \left[\frac{C_c}{(1-\eta_O)^2}+\frac{C_h}{(1-\eta_C)^2}\right],
 \end{split}
\end{gather}
where, $T_c$ ($T_h$) is the temperature of the cold (hot) reservoir, $k_B$ is the Boltzmann constant, $\eta_O=1-\epsilon$, $\sigma_c\equiv E_c/T_c$ ($\sigma_h\equiv E_h/T_h$), $E_c$ ($E_h$) is the internal energy of the working substance corresponding to the equilibrium state at $t=0$ ($t=\tau_c+t_h$), and $C_c\equiv\partial E_c/\partial T_c$ ($C_h\equiv\partial E_h/\partial T_h$) is the heat capacity at constant volume. In addition, the the average heat absorbed from the hot reservoir and the average entropy production of the cycle follow as $\left\langle q\right\rangle=-\left\langle w\right\rangle/\eta_O$  and $\left\langle \delta s\right\rangle=\beta_c\left\langle q\right\rangle(\eta_C-\eta_O)$, respectively. If the working substance statisfies the equipartition theorem $E_c\propto k_B T_c$ and $E_h\propto k_B T_h$ in the high-temperature limit, one has
\begin{equation}
\label{e13}
\frac{\mathrm{\Delta}\eta^{2}}{\left\langle \eta\right\rangle ^{2}}=\frac{\mathrm{\Delta}w^{2}}{\left\langle w\right\rangle ^{2}}=\frac{1}{\left\langle \delta s\right\rangle}\left(\frac{1-\eta_O}{1-\eta_C}+\frac{1-\eta_C}{1-\eta_O}\right)\geq\frac{2}{\left\langle \delta s\right\rangle}
\end{equation}
with the equal sign saturated at $\eta_O=\eta_C$. The inequality~(\ref{e13}) is consistent with the TUR in steady states~\cite{barato2015,pietz2016} or in a specific Otto cycle~\cite{lee2021}.

Moreover, due to the third law of thermodynamics, $\left\langle \delta s\right\rangle\to 0$ in the low-temperature limit. Using the property of the function $f(x)$ in Eq.~(\ref{eq:11}), the TUR for the efficiency is also reproduced in the low-temperature limit. Consequently, we expect that  the TUR for efficiency is valid for an arbitrary temperature under these conditions. For a finite-time cycle, we numerically study the TUR in a specific model below.
%Hence, the statistics of the efficiency of the Otto cycle satisfies the thermodynamic uncertainty relation in both the high-temperature limit (Eq.~(\ref{e13})) and the low-temperature limit.

\section{Quantum harmonic-oscillator heat engine}
\label{s4}

In this section, we illustrate our general discussions above with
a specific example: a quantum harmonic-oscillator being the working substance of the Otto cycle.
The frequency is changed from $\omega_{0}$ to $\omega_{1}$ ($\omega_{1}>\omega_{0}$) in the adiabatic compression process. Then, according to  Refs.~\cite{no2008, zy2019}, $\chi_{c}\left(u,v\right)$ and $\chi_{h}\left(u,v\right)$ of the joint characteristic function
in Eq.~(\ref{eq:12}) are explicitly obtained as (see Appendix C for detailed derivation)

%\begin{equation}
%\chi\left(u,v\right)=\chi_{c}\left(u,v\right)\chi_{h}\left(u,v\right),\label{eq:distribution}
%\end{equation}
%where

\begin{widetext}
\begin{equation}
\label{eq:distribution}
\chi_{c}\left(u,v\right)=2\sinh\left(\frac{\beta_{c}\omega_{0}}{2}\right)\left\{ 2\cos\left[\left(u-v\right)\omega_{1}\right]\cos\left[\left(u-i\beta_{c}\right)\omega_{0}\right]+2Q_{c}\sin\left[\left(u-v\right)\omega_{1}\right]\sin\left[\left(u-i\beta_{c}\right)\omega_{0}\right]-2\right\} ^{-\frac{1}{2}},
\end{equation}
and

\begin{equation}
\label{eq:distribution2}
\chi_{h}\left(u,v\right)=2\sinh\left(\frac{\beta_{h}\omega_{1}}{2}\right)\left\{ 2\cos\left(u\omega_{0}\right)\cos\left[\left(u-v-i\beta_{h}\right)\omega_{1}\right]+2Q_{h}\sin\left(u\omega_{0}\right)\sin\left[\left(u-v-i\beta_{h}\right)\omega_{1}\right]-2\right\} ^{-\frac{1}{2}},
\end{equation}
\end{widetext}
where $Q_{c(h)}\geq1$ is the corresponding non-adiabatic factor~\cite{jaramillo2019,husimi1953}.
The equal sign is hold when the quantum adiabatic condition
is satisfied. In the following, we study the efficiency statistics of the Otto cycle in different circumstances.

\subsection{Average efficiency and efficiency distribution of the heat engine}

\begin{figure}
\begin{centering}
\includegraphics[width=8.5cm]{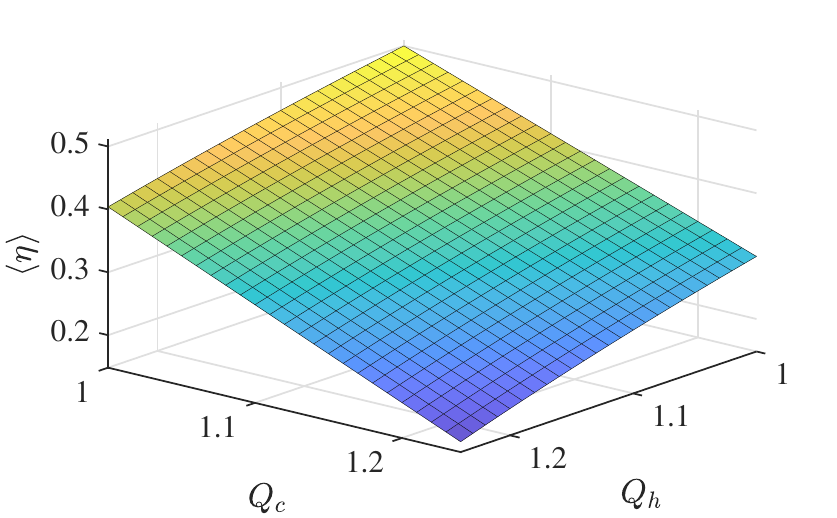}
\par\end{centering}
\caption{\label{fig:Average efficiency}Average efficiency as the function of
$Q_{c}$ and $Q_{h}$. In this figure, we use $\eta_{O}=0.5$, $T_{h}=1$
$\eta_{C}=0.8$, and $\omega_{1}=1$.}
\end{figure}

The average output work and average absorbed heat per cycle can be
obtained using Eqs.~(\ref{aw}),(\ref{aq}),(\ref{eq:distribution}),(\ref{eq:distribution2})
as

\begin{align}
-\left\langle w\right\rangle =-\frac{1}{2}\left[\left(\omega_{1}Q_{c}-\omega_{0}\right)\vartheta_{c}-\left(\omega_{1}-\omega_{0}Q_{h}\right)\vartheta_{h}\right],\label{eq:work}
\end{align}
and

\begin{equation}
\left\langle q\right\rangle =\frac{\omega_{1}}{2}\left(\vartheta_{h}-Q_{c}\vartheta_{c}\right),\label{eq:heat}
\end{equation}
where

\begin{equation}
\vartheta_{h}\equiv\coth\frac{\beta_{h}\omega_{1}}{2},\vartheta_{c}\equiv\coth\frac{\beta_{c}\omega_{0}}{2}.
\end{equation}
Substituting Eqs.~(\ref{eq:work}) and (\ref{eq:heat}) into Eq.~(\ref{eq:3}), the average efficiency is obtained as

\begin{equation}
\label{e20}
\left\langle \eta\right\rangle =\frac{\left(\omega_{1}-\omega_{0}Q_{h}\right)\vartheta_{h}-\left(\omega_{1}Q_{c}-\omega_{0}\right)\vartheta_{c}}{\omega_{1}\left(\vartheta_{h}-Q_{c}\vartheta_{c}\right)}
\end{equation}
In the quasistatic limit, the non-adiabatic factors $Q_{c,h}=1$, and
the average efficiency is the Otto efficiency $\eta_{O}=1-\epsilon$ ($\epsilon=\omega_{0}/\omega_{1}$).
%\begin{equation}
%\left\langle \eta\right\rangle =1-\frac{\omega_{0}}{\omega_{1}}=1-\epsilon=\eta_{O},
%\end{equation}
Then, Eq.~(\ref{e20}) is further expressed with $\eta_{O}$
as

\begin{equation}
\left\langle \eta\right\rangle =\eta_{O}-\frac{\epsilon\sum_{\alpha=h,c}\vartheta_{\alpha}\left(Q_{\alpha}-1\right)}{\vartheta_{h}-Q_{c}\vartheta_{c}}.\label{eq:average-eta}
\end{equation}
This result means that the non-adiabatic effect  decreases the
average efficiency of the engine, which is demonstrated in Fig.~\ref{fig:Average efficiency}.

On the other hand, the variance of the efficiency is $\Delta\eta^{2}=\Delta w^{2}/\left\langle q\right\rangle ^{2}$,
%\begin{equation}
%\Delta\eta^{2}\equiv\left\langle \eta^{2}\right\rangle -\left\langle \eta\right\rangle ^{2}=\frac{\Delta w^{2}}{\left\langle q\right\rangle ^{2}},\label{eq:efficiency-fluctuation}
%\end{equation}
where the variance of  work  is obtained from Eq.~(\ref{dw})
as
\begin{widetext}
\begin{equation}
\Delta w^{2}=\frac{\omega_{1}^{2}}{4}\left(1-\epsilon\right)^{2}\left(\vartheta_{h}^{2}+\vartheta_{c}^{2}-2\right)+\frac{\omega_{1}^{2}}{2}\left[\vartheta_{h}^{2}\left(Q_{h}^{2}-1\right)\epsilon^{2}+\vartheta_{c}^{2}\left(Q_{c}^{2}-1\right)-\epsilon\sum_{\alpha=h,c}\left(Q_{\alpha}-1\right)\left(\vartheta_{\alpha}^{2}-1\right)\right],
\end{equation}
\end{widetext}
%\begin{equation}
%\Delta w_{adi}^{2}=\frac{\omega_{1}^{2}}{4}\left(1-\epsilon\right)^{2}\left(\vartheta_{h}^{2}+\vartheta_{c}^{2}-2\right),
%\end{equation}
%is the work fluctuation for the quantum adiabatic driving.
In the quasistatic limit, the variance of  efficiency
accordingly becomes

\begin{equation}
\label{eq:efficiency-fluctuation}
\Delta\eta_{adi}^{2}=\frac{\left(1-\epsilon\right)^{2}\left(\vartheta_{h}^{2}+\vartheta_{c}^{2}-2\right)}{\left(\vartheta_{h}-\vartheta_{c}\right)^{2}}.
\end{equation}
It is worth mentioning that the efficiency fluctuation does not
vanish for a quantum harmonic-oscillator Otto cycle in the quasistatic limit, while
the fluctuation of the previous version of the stochastic efficiency vanishes in this case~\cite{denzler2020}.

\begin{figure}
\begin{raggedright}
(a)
\par\end{raggedright}
\begin{raggedright}
\includegraphics[width=8.5cm]{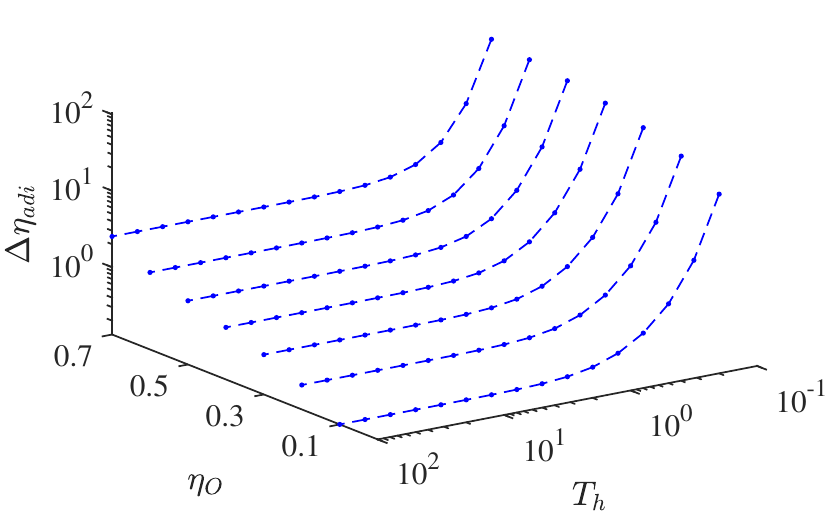}
\par\end{raggedright}
\begin{raggedright}
(b)
\par\end{raggedright}
\begin{raggedright}
\includegraphics[width=8.5cm]{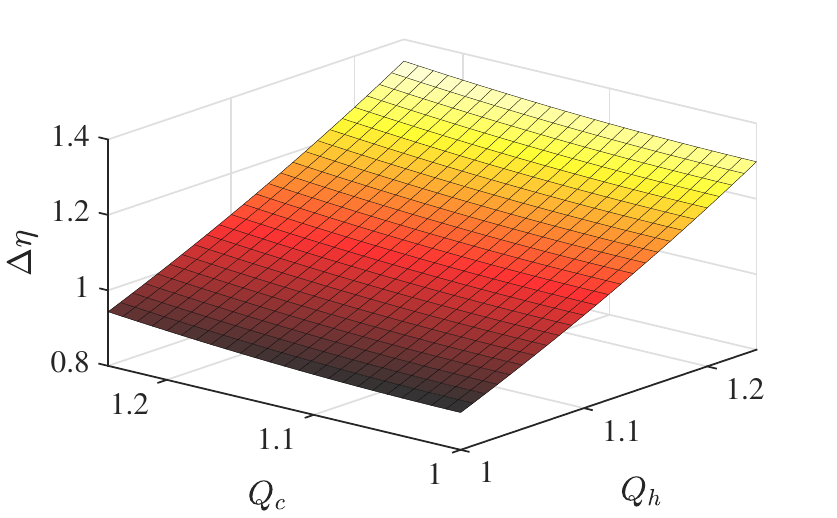}
\par\end{raggedright}

\caption{\label{fig:Efficiency-fluctuation.-(a)}Efficiency fluctuation. (a)
Efficiency fluctuation as the function of temperature $T_{h}$ with
different $\eta_{O}$ in the quasistatic limit. (b) Efficiency fluctuation
as the function of $Q_{c}$ and $Q_{h}$, where $\eta_{O}=0.5$ and
$T_{h}=1$ are chosen. In this figure, $\eta_{C}=0.8$ and $\omega_{1}=1$
are fixed.}
\end{figure}

It is shown in Fig.~\ref{fig:Efficiency-fluctuation.-(a)} (a) that,
in the quasistatic limit ($Q_{c,h}=1$), the efficiency fluctuation
decreases as the temperature increases, which is consistent with the
TUR for efficiency (Eq.~(\ref{e13})) since
$\left\langle \delta s\right\rangle $ increases with the temperature
increases. In the non-adiabatic case, the efficiency fluctuation as
the function of $Q_{c}$ and $Q_{h}$ is illustrated in Fig.~\ref{fig:Efficiency-fluctuation.-(a)}
(b),
which reflects the enhancement of the fluctuation due
to the non-adiabatic driving.

\begin{figure}
\begin{raggedright}
(a)
\par\end{raggedright}
\begin{raggedright}
\includegraphics[width=8.5cm]{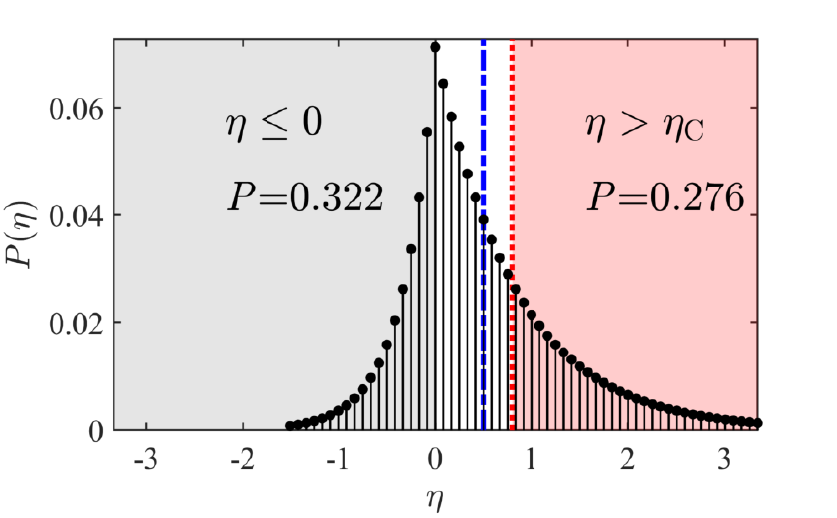}
\par\end{raggedright}
\begin{raggedright}
(b)
\par\end{raggedright}
\begin{raggedright}
\includegraphics[width=8.5cm]{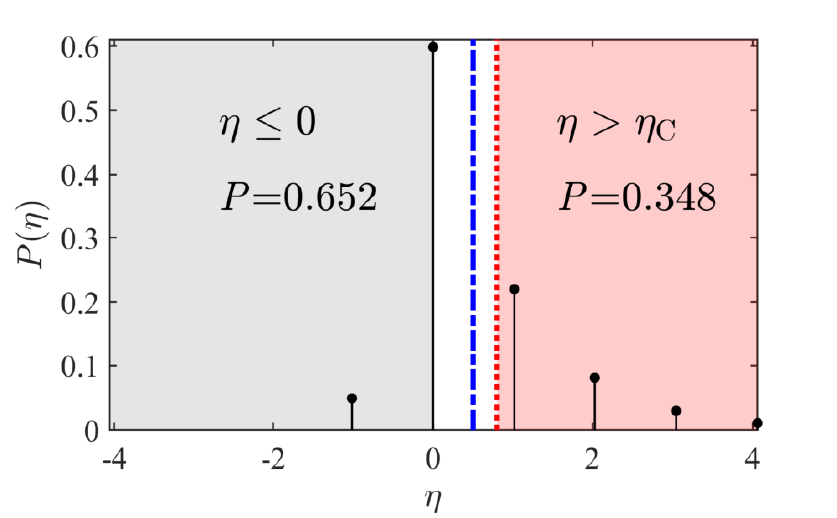}
\par\end{raggedright}

\caption{\label{fig:Efficiency-distributionA.-In}Efficiency distribution in adiabatic case with $Q_{h}=Q_{c}=1$. The probability distribution of efficiency is plotted with the black dots. The Otto efficiency and Carnot efficiency are respectively represented with the blue dash-dotted line and the red dotted line. The (gray) area in the left side denotes the negative
work output regime; the (light red) area in the right side of the
red dotted line represents the regime of $\eta>\eta_{C}$. The parameters
are chosen as: (a) $T_{h}=10$, $T_{c}=2$; (b) $T_{h}=1$, $T_{c}=0.2$. In this figure, we choose $\omega_{0}=0.5$, $\omega_{1}=1$, and Carnot efficiency
is fixed at $0.8$.}
\end{figure}

\begin{figure}
\begin{raggedright}
(a)
\par\end{raggedright}
\begin{raggedright}
\includegraphics[width=8.5cm]{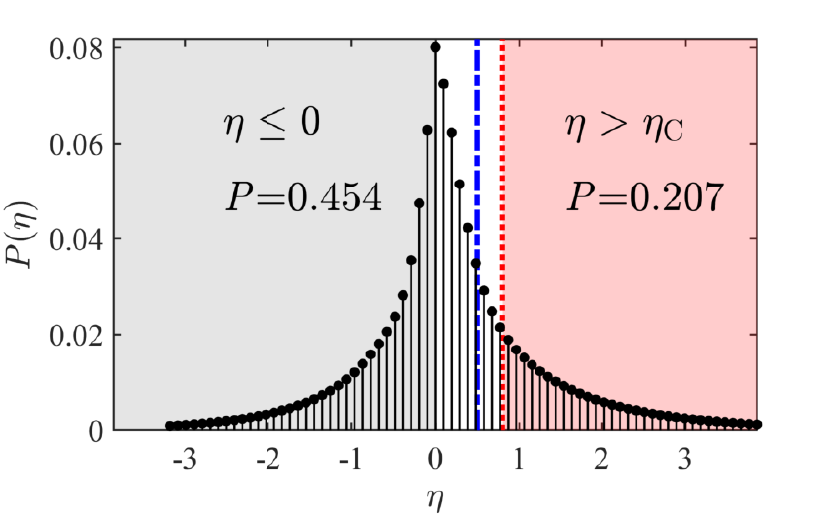}
\par\end{raggedright}
\begin{raggedright}
(b)
\par\end{raggedright}
\begin{raggedright}
\includegraphics[width=8.5cm]{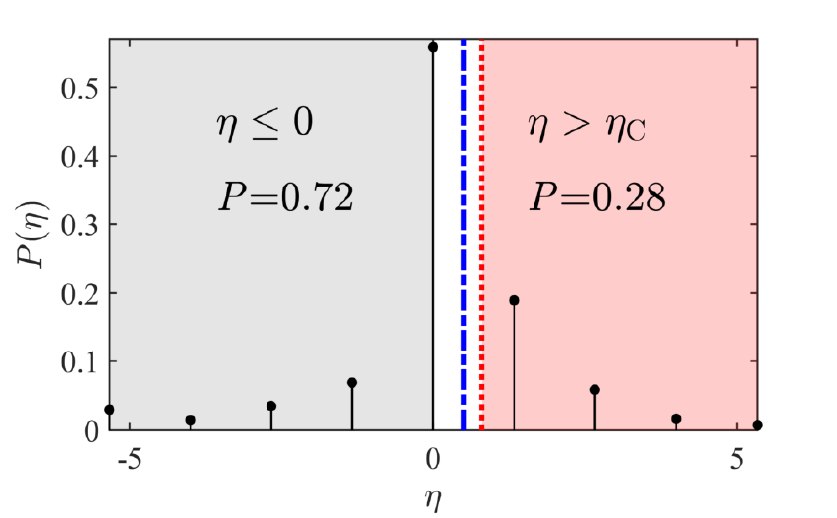}
\par\end{raggedright}

\caption{\label{fig:Efficiency-distributionNA.-In}Efficiency distribution in non-adiabatic case with $Q_{h}=Q_{c}=1.2$. The probability distribution of efficiency is plotted
with the black dots. The Otto efficiency and Carnot efficiency are
respectively represented with the blue dash-dotted line and the red
dotted line. The (gray) area in the left side denotes the negative
work output regime; the (light red) area in the right side of the
red dotted line represents the regime of $\eta>\eta_{C}$. The parameters
are chosen as: (a)$T_{h}=10$, $T_{c}=2$; (b) $T_{h}=1$, $T_{c}=0.2$. In this figure, we
choose $\omega_{0}=0.5$, $\omega_{1}=1$, and Carnot efficiency
is fixed at $0.8$.}
\end{figure}

The efficiency distribution is obtained with $\chi\left(u,v\right)$
by the discrete Fourier transform. With different chosen parameters,
we plot the efficiency distribution in Fig.~\ref{fig:Efficiency-distributionA.-In} (adiabatic case) and Fig.~\ref{fig:Efficiency-distributionNA.-In} (non-adiabatic case)
with the black dots. As comparisons, the Otto efficiency and Carnot
efficiency are respectively represented with the blue dash-dotted
line and the red dotted line. %It is illustrated in the figures that, according to the stochastic efficiency we defined,
And we show the probability that the efficiency of a stochastic Otto cycle surpasses the Carnot efficiency in the figure. In addition, %by comparing Fig.~\ref{fig:Efficiency-distributionA.-In} (a) and  Fig.~\ref{fig:Efficiency-distributionA.-In} (b) or Fig.~\ref{fig:Efficiency-distributionNA.-In} (a) and  Fig.~\ref{fig:Efficiency-distributionNA.-In} (b)
one can infer  that lower temperature leads to greater probability of the heat engine surpassing the Carnot efficiency. Meanwhile, the lower temperature
increases the probability of the engine to be useless, namely, the
engine outputs negative work.

%\begin{figure}
%\begin{centering}
%\subfloat[]{\includegraphics[width=8.5cm]{Figure/probabilityOtto-Th=0\lyxdot 2}}\subfloat[]{\includegraphics[width=8.5cm]{Figure/probabilityCarnot-Th=0\lyxdot 2}}
%\par\end{centering}
%\begin{centering}
%\subfloat[]{\includegraphics[width=8.5cm]{Figure/probabilityOtto-Th=1}}\subfloat[]{\includegraphics[width=8.5cm]{Figure/probabilityCarnot-Th=1}}
%\par\end{centering}
%\begin{centering}
%\subfloat[]{\includegraphics[width=8.5cm]{Figure/probabilityOtto-Th=5}}\subfloat[]{\includegraphics[width=8.5cm]{Figure/probabilityCarnot-Th=5}}
%\par\end{centering}
%\centering{}\subfloat[]{\includegraphics[width=8.5cm]{Figure/probabilityOtto-Th=50}}\subfloat[]{\includegraphics[width=8.5cm]{Figure/probabilityCarnot-Th=50}}\caption{\label{fig:Efficiency-distribution.-In-1}The probabilities for the
%heat engine to surpass Otto efficiency and Carnot efficiency as the
%function of $Q_{h}$ and $Q_{c}$ . In this figure, (a), (c), (e),
%(g) represent the probabilities that the heat engine efficiency surpass
%the Otto efficiency with different temperature. (b), (d), (f), (h)
%represent the probabilities that the heat engine efficiency surpass
%the Carnot efficiency with different temperature. (a), (b)$T_{h}=0.2$.
%(c), (d) $T_{h}=1$. (e), (f) $T_{h}=10$. (g), (h) $T_{h}=50$. The
%initial and final frequencies of the harmonic oscillator in the adiabatic
%process are fixed at $\omega_{0}=0.5$ and $\omega_{1}=1$. The Carnot
%efficiency is fixed at $\eta_{C}=0.8$}
%\end{figure}

\subsection{Finite-time performance of the heat engine}

\begin{figure}
\centering{}\includegraphics[width=8.5cm]{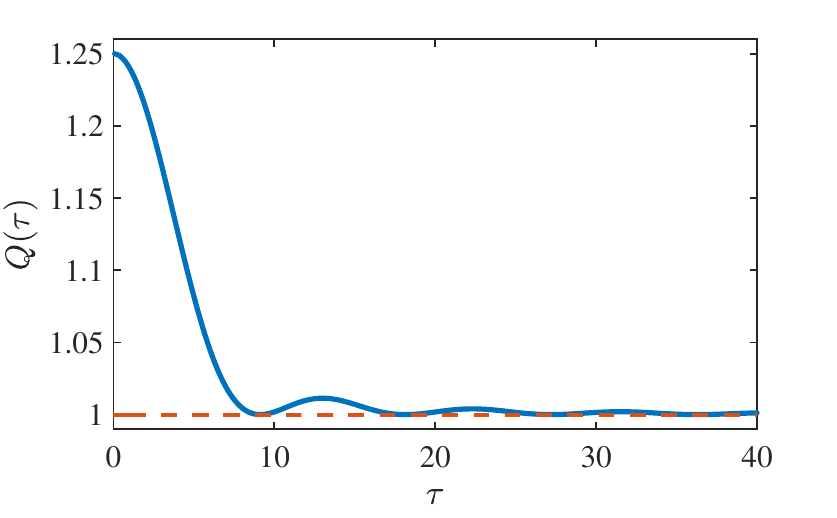}\caption{\label{fig:Time-dependence-of-the}Time dependence of the non-adiabatic
factor. In this figure, the blue solid curve represents $Q(\tau)$
in Eq. (\ref{eq:Q(t)}), the orange dashed line is $Q(\tau)=1$. The
initial and final frequencies of the harmonic oscillator in the adiabatic
process are chosen as $\omega_{0}=0.5$ and $\omega_{1}=1$.}
\end{figure}

To further explore the finite-time performance of the cycle, we
first analyze the explicit time dependence of the non-adiabatic factors for a specific protocol.
%namely, $Q_{h(c)}\equiv Q(\tau_{h(c)})$, where $\tau_{h(c)}$ is
%the operation time of the finite-time adiabatic process. Generally,
%the non-adiabatic factor depends on the specific dynamic of the system
%as well as the protocol to tune the system.
For an adiabatic process with frequency changed from $\omega_i$ to $\omega_f$ during time $t\in[0,\tau]$,
the time dependence of the frequency of the harmonic oscillator is~\cite{Chen20192,beau2016,lee2021}

\begin{equation}
\label{protocol}
\omega(t)=\frac{\omega_{i}}{(\omega_{i}/\omega_{f}-1)t/\tau+1}.
\end{equation}
Then, the non-adiabatic factor $Q(\tau)$ is obtained as (See Appendix D
for detailed derivation)
\begin{equation}
Q(\tau)=1+\frac{1-\cos\left[\sqrt{a^{2}\tau^2-1}\ln(\omega_f/\omega_i)\right]}{a^{2}\tau^2-1},\label{eq:Qtau}
\end{equation}
where
\begin{equation}
a\equiv\frac{2\omega_{f}\omega_{i}}{\omega_{f}-\omega_{i}}.
\end{equation}
As shown in Fig.~\ref{fig:Time-dependence-of-the}, the non-adiabatic
factor $Q(\tau)$ (blue solid line) oscillates with the driving
time $\tau$, reflecting the quantum coherence effect in the non-adiabatic
transition. The orange dashed line represents $Q(\tau)=1$, which is achieved
for the quantum adiabatic driving or with some special values of $\tau$~\cite{beau2016,Chen20192}.

In the following, we adopt the protocol of Eq. (\ref{protocol}) for the finite-time adiabatic processes in the Otto cycle, then we use of the explicit form of $Q(\tau)$ given in Eq. (\ref{eq:Qtau}) to study the power at maximum efficiency (PME) and efficiency at maximum power (EMP)
of the cycle. In this sense, the non-adiabatic factors become $Q_{c(h)}=Q(\tau_{c(h)})$, and then the average power $\left\langle P(\tau_{c},\tau_{h})\right\rangle \equiv-\left\langle w\right\rangle /(\tau_{h}+\tau_{c})$ and the efficiency
$\left\langle \eta(\tau_{c},\tau_{h})\right\rangle$ of the Otto engine are respectively

\begin{equation}
\left\langle P(\tau_{c},\tau_{h})\right\rangle=\frac{\omega_{1}}{2}\frac{\left\{ \left[Q(\tau_{c})-\epsilon\right]\vartheta_{c}-\left[1-\epsilon Q(\tau_{h})\right]\vartheta_{h}\right\} }{\tau_{h}+\tau_{c}},\label{eq:Power}
\end{equation}
and

\begin{equation}
\left\langle \eta(\tau_{c},\tau_{h})\right\rangle =\eta_{O}-\frac{\epsilon\sum_{\alpha=h,c}\vartheta_{\alpha}\left[Q(\tau_{\alpha})-1\right]}{\vartheta_{h}-Q(\tau_{c})\vartheta_{c}},\label{eq:efficiency}
\end{equation}
where the total duration of the two isochoric processes, i.e., $t_c+t_h$, is assumed to be
much smaller than $\tau_c+\tau_h$, and is thus ignored.

%In Fig.~\ref{fig:Power-efficiency-trade-off-of},
%\begin{figure}
%\centering{}\includegraphics[width=8.5cm]{Figure/P-eta-deta}\caption{\label{fig:Power-efficiency-trade-off-of}the efficiency fluctuation as a function of average power and average efficiency of the Otto cycle. We choose $\omega_{0}=0.5$, $\omega_{1}=1$, and
%$\tau_{c(h)}\in\left[0.1,200\right]$ to plot $\left\langle P(\tau_{c},\tau_{h})\right\rangle $,
%$\left\langle \eta(\tau_{c},\tau_{h})\right\rangle $ and $\Delta\eta(\tau_{c},\tau_{h})$.
%%The initial and final frequencies of the harmonic oscillator in the
%%adiabatic process is chosen as $\omega_{0}=0.5$ and $\omega_{1}=1$.
%The Otto efficiency in the figure is $\eta_{O}=0.5$. The (green) circle,
%triangle, and square respectively denote the largest, second largest,
%third largest value of $P(\tau_{c}^{*},\tau_{h}^{*})$ in Eq. (\ref{eq:P-etaotto}).}
%\end{figure}
%we plot the efficiency fluctuation as the function average power
%and average efficiency of for such Otto cycle with different $(\tau_{c},\tau_{h})$.
%In this figure, the efficiency fluctuation is proportional to the
%brightness in the corresponding area. The boundary of the shaded area
%implies the trade-off between power and efficiency of the engine~\cite{Chen20192}.
%It is shown in the figure that the efficiency fluctuation decreases
%as the average efficiency increases. The efficiency fluctuation reaches
%it maximum in the lower-left corner of the figure, where average power
%and average efficiency are both small.

Since $Q_{c(h)}=1$ can be achieved within finite time, the average efficiency of some cycles approach the Otto efficiency
$\eta_{O}$ with non-vanishing power. These cycles happen to have the
special operation time sets $(\tau_{c}^{*},\tau_{h}^{*})$ corresponding
to $Q_{h}(\tau_{h}^{*})=Q_{c}(\tau_{c}^{*})=1$. %, as illustrated in
%Fig.~\ref{fig:Time-dependence-of-the}.
With the help of Eq.~(\ref{eq:Qtau}),
one finds the special operation time follows as

\begin{equation}
\sqrt{a^{2}\tau^2-1}\ln\epsilon=2n\pi,n=1,2,3...,
\end{equation}
namely,

\begin{equation}
\tau_{h,c}^{*}=\frac{\eta_{O}}{\omega_{0}}\sqrt{\frac{1}{4}+\left(\frac{n_{h,c}\pi}{\ln\epsilon}\right)^{2}}.
\end{equation}
Therefore, the PME is

\begin{equation}
P(\tau_{c}^{*},\tau_{h}^{*})=\frac{\omega_{0}\omega_{1}\left(\vartheta_{h}-\vartheta_{c}\right)}{2\sum_{\alpha=h,c}\sqrt{1/4+\left(n_{\alpha}\pi/\ln\epsilon\right)^{2}}}.\label{eq:P-etaotto}
\end{equation}
It should be noted that in the usual finite-time thermodynamic cycles, the PME generally approaches zero ~\cite{Shiraishi2016, Ma2018, Chen2019,Yuan2021}. Here, thanks to the special protocol we have chosen to realize the quantum adiabatic process in finite time, the current quantum Otto cycle outputs non-zero or even relatively large power (comparable to the maximum power) when the Otto efficiency is reached. Obviously, the maximum $P(\tau_{c}^{*},\tau_{h}^{*})$ is

\begin{equation}
P_{max}(\tau_{c}^{*},\tau_{h}^{*})=\frac{\omega_{0}\omega_{1}\left(\vartheta_{h}-\vartheta_{c}\right)}{2\sqrt{1+\left(2\pi/\ln\epsilon\right)^{2}}},
\end{equation}
which is achieved at $n_{c}=n_{h}=1$. Besides,
the second largest and third largest
power are reached at ($n_{c}=1,n_{h}=2$) and ($n_{c}=2,n_{h}=2$),
respectively. For $(n_{\alpha}\pi/\ln\epsilon)^{2}\gg1/4$, Eq. (\ref{eq:P-etaotto})
can be approximated as

\begin{equation}
P(\tau_{c}^{*},\tau_{h}^{*})\approx\frac{\omega_{0}\omega_{1}\ln\epsilon\left(\vartheta_{h}-\vartheta_{c}\right)}{2\pi\left(n_{c}+n_{h}\right)},
\end{equation}
which shows that $P(\tau_{c}^{*},\tau_{h}^{*})$ is a monotonically decreasing quasi-continuous function of $n_c$ and $n_h$. %(the left side of the green square in Fig.~\ref{fig:Power-efficiency-trade-off-of}).
%This fact in shown Fig.~\ref{fig:Power-efficiency-trade-off-of}
%that on the left side of the green square, the power-efficiency trade-off
%tending to the black line are almost continuous.
\begin{figure}
\includegraphics[width=8.5cm]{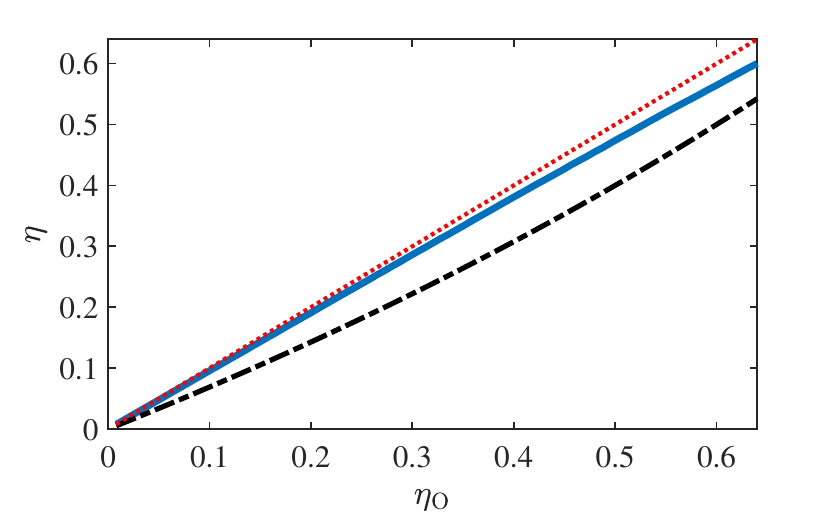}\caption{\label{fig:Efficiency-at-maximum}Efficiency at maximum power
of the Otto engine as the function of $\eta_O$.  In this figure, $\eta_O/\eta_{\mathrm{C}}=0.8$ is fixed. The blue
solid curve represents the EMP of the Otto
engine. The black dash-dotted curve is the upper bound for EMP, $\eta_{+}=2\eta_{O}/(3-\eta_{O})$, of the Otto
cycle obtained in Ref.~\cite{Chen2019} without considering the oscillation of work. The red dotted line denotes the Otto efficiency.}
\end{figure}

In addition, the EMP of this Otto engine as
the function of $\eta_O$ is illustrated in Fig. \ref{fig:Efficiency-at-maximum}.
As shown in this figure, the EMP of our cycle (blue solid curve) is found to surpass the upper bound, $\eta_{+}=2\eta_{O}/(3-\eta_{O})$ (black dashed curve), of the Otto cycle's EMP
without considering the oscillation of the output work~\cite{Chen2019}. This
indicates that the oscillation of the output work (due to quantum coherence) are conducive to improving the
EMP. %In this example, we fix the ratio of adiabatic Otto efficiency
%(red dotted curve) to Carnot efficiency to $0.8$, namely, $\eta_{O}/\eta_{\mathrm{C}}=0.8$.

\begin{figure}
\begin{raggedright}
(a)
\par\end{raggedright}
\begin{raggedright}
\includegraphics[width=8.5cm]{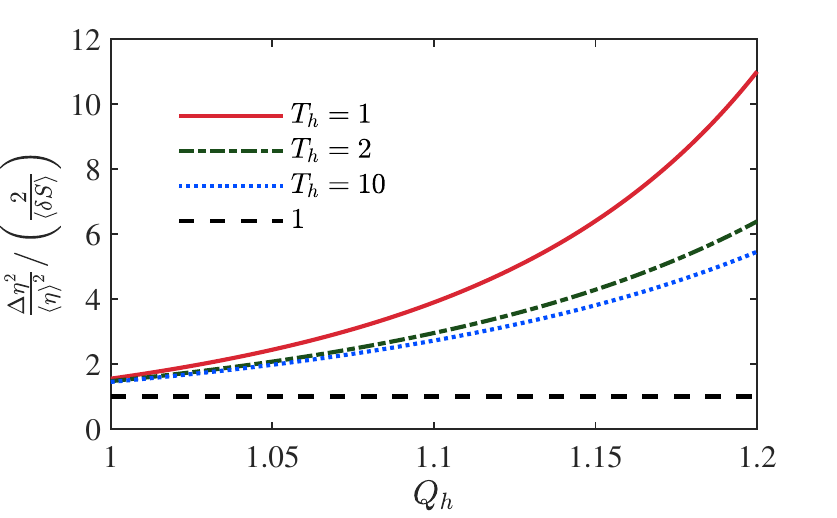}
\par\end{raggedright}
\begin{raggedright}
(b)
\par\end{raggedright}
\begin{raggedright}
\includegraphics[width=8.5cm]{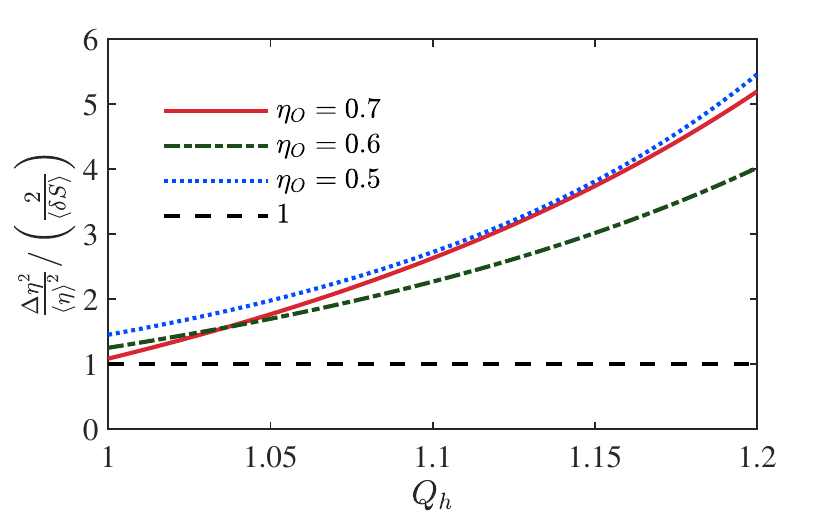}
\par\end{raggedright}

\caption{Thermodynamic uncertainty relation for the efficiency with different non-adiabatic
factors. (a) $T_{h}=1,2,10$, $\eta_{O}=0.5$; (b) $T_{h}=10$, $\eta_{O}=0.5,0.6,0.7$.
In this figure, $Q_{c}=1$, $\eta_{C}=0.8$ and $\omega_{1}=1$ are
fixed parameters.}\label{fig:tur}

\end{figure}
\subsection{Thermodynamic uncertainty relation (TUR) for efficiency}

Because the spectra of a quantum harmonic oscillator have scale property and the system follows the equipartition theorem in the high-temperature limit, we conclude that in the quasistatic limit, the TUR (Eq.~(\ref{e13})) is valid according to the discussions in Sec.~\ref{s3b}.

For the non-adiabatic driving cycle, the results are shown in Fig.~\ref{fig:tur}. The TUR (Eq.~(\ref{e13})) is still valid since $\frac{\Delta\eta^{2}}{\left\langle \eta\right\rangle ^{2}}/\left(\frac{2}{\left\langle \delta S\right\rangle }\right)$
increases monotonically with $Q_{h}$ and $Q_{c}$. Here, without loss of generality, we take $Q_h$ as the independent variable in the figure. On the contrary, the TUR may be violated due to the incomplete thermalization in the isochoric processes~\cite{lee2021}. One can conclude from Fig.~\ref{fig:tur}(a) that higher temperature makes $\frac{\Delta\eta^{2}}{\left\langle \eta\right\rangle ^{2}}/\left(\frac{2}{\left\langle \delta S\right\rangle }\right)$
lower. Moreover, as shown in Fig.~\ref{fig:tur}(b), when $Q_h\rightarrow1$, $\frac{\Delta\eta^{2}}{\left\langle \eta\right\rangle ^{2}}/\left(\frac{2}{\left\langle \delta S\right\rangle }\right)$ is closer to $1$ in the case with $\eta_{O}=0.7$. This is consistent
with the discussions in  Sec.~\ref{s3b} that the condition for $\frac{\Delta\eta^{2}}{\left\langle \eta\right\rangle ^{2}}/\left(\frac{2}{\left\langle \delta S\right\rangle }\right)\to1$ is
$\eta_{O}\rightarrow\eta_{C}$.

\section{Summary and discussion}
\label{s5}

In this paper, we come up with
a new definition of the stochastic efficiency for heat engine in micro scale. The moments of the efficiency always exist,
and its mean
value is equal to the conventional efficiency.
Moreover, the fluctuation theorems are resexpressed via the efficiency. For spectra of the working substance with scale property, the statistics of the efficiency is fully determined by the partition functions of the working substance in the quasistatic limit. Importantly, we reveal the connection between the TUR and the equipartition theorem.

For a quantum Otto cycle with a harmonic oscillator
being the working substance, we obtain the exact expression of the
joint characteristic function of   work and heat. We find that the Otto efficiency can be reached with a finite output power (the power at maximum efficiency) with some special duration  and the EMP surpasses the upper bound obtained in Ref.~\cite{Chen2019}.

%It is worth mentioning that the
%definition in Eq. (\ref{eq:3}) is not the unique definition of the
%stochastic efficiency which satisfies the requirements above. For
%other definitions, the fluctuation of the stochastic efficiency may
%not be determined by the output work but by a mixture of the output
%work and the absorbed heat. In those cases, the fluctuation of the efficiency characterizes other statistical properties of the cycle.
The theoretical predictions of current study can be tested on some state-of-art experiments, such as the Brownian particle system~\cite{martinez2016} and trapped ion system~\cite{Singleatom2012}. As a direct extention, similarly to the stochastic efficiency defined here, the coefficient of performance of a refrigerator can be defined as the ratio of the stochastic released heat to the average input work. Then, the statistics of a stochastic refrigerator can be further discussed. Besides, it is expected that the many-body effect of the working substance~\cite{beau2016, jaramillo2019, beng2018, chen2018, Ma2017, zy2020} and the influences of the control protocols for the cycle~\cite{Ma20182,Ma2020,Brandner2020} on the efficiency statistics and TUR will be taken into consideration in future investigations.

\begin{acknowledgments}
This work is supported by the National Natural Science Foundation
of China (NSFC) (Grants No. 11534002, No. 11875049, No. U1730449,
No. U1530401, and No. U1930403), and the National Basic Research Program
of China (Grants No. 2016YFA0301201). Y. H. Ma is supported by the China Postdoctoral Science Foundation (Grant No. BX2021030).
\end{acknowledgments}

 \renewcommand{\theequation}{A\arabic{equation}}

 \setcounter{equation}{0}

\section*{Appendix A: Proof of $\langle \eta \rangle \leq\eta_O$}

As a result of the minimum work principle~\cite{alla2005}, when the energy levels do not cross during the driving, the average work under finite-time driving $\left\langle w_c(e)\right\rangle$ is not less than it under quantum adiabatic driving $\left\langle w_c(e)\right\rangle_{adi}$. Namely, $\delta w_{c(e)}\equiv \left\langle w_{c(e)}\right\rangle-\left\langle w_{c(e)}\right\rangle_{adi}\geq 0$.
Thus, the average efficiency of a quantum Otto cycle with the complete thermalization satisfies
%The efficiency of a finite-time Otto cycle with the complete thermalization
%is
\begin{gather}
 \begin{split}
\langle\eta\rangle&=-\frac{\left\langle w\right\rangle}{\left\langle q\right\rangle}=-\frac{\left\langle w_c\right\rangle _{adi}+\left\langle w_e\right\rangle _{adi}+\delta w_{c}+\delta w_{e}}{\left\langle q\right\rangle _{adi}-\delta w_{c}}\\
&\leq-\frac{\left\langle w_c\right\rangle _{adi}+\left\langle w_e\right\rangle _{adi}}{\left\langle q\right\rangle _{adi}}=\eta_O,
 \end{split}
\end{gather}
for $\left\langle q\right\rangle _{adi}>-\left\langle w_c\right\rangle _{adi}-\left\langle w_e\right\rangle _{adi}>0$, and $\delta w_{c}\geq0, \delta w_{e}\geq0$, where  $\left\langle q\right\rangle _{adi}$ denotes  the heat absorbed from the hot reservoir in the quasistatic limit.
%From the fact that $\left\langle q\right\rangle _{\mathrm{quasi}}>-\left\langle w\right\rangle _{\mathrm{quasi}}>0$,
%$\delta w_{c}(\tau_{c})\geq0$ and $\delta w_{e}(\tau_{h})\geq0$,
%we obtain
%
%\begin{equation}
%\eta\leq-\frac{\left\langle w\right\rangle _{\mathrm{quasi}}}{\left\langle q\right\rangle _{\mathrm{quasi}}}=\eta_{O}.
%\end{equation}

 \renewcommand{\theequation}{B\arabic{equation}}

 \setcounter{equation}{0}

 \section*{Appendix B: The joint characteristic function for a quantum Otto cycle in the quasistatic limit}

According to Eq.~(\ref{eq:12}), the expression of the joint characteristic
function $\chi(u,v)$ is obtained by a transformation of the characteristic
function of work $\chi_{c}(u)\equiv\chi_{c}(u,0)$ and $\chi_{h}(u)\equiv\chi_{h}(u,0)$, i.e.,
\begin{gather}
 \begin{split}
 \label{b1}
\chi_{c}(u,v)&=\chi_{c}(u)|_{u\rightarrow u-v,\beta_{c}\rightarrow\beta_{c}+iv}\\
\chi_{h}(u,v)&=\chi_{h}(u)|_{\beta_{h}\rightarrow\beta_{h}-iv}.
 \end{split}
\end{gather}
with scale property ($E_n^1=E_n^0/\epsilon$), in the quasistatic limit, the expressions of the characteristic function $\chi_c(u)$ and $\chi_h(u)$ read
\begin{gather}
 \begin{split}
\chi_c(u)&=\sum_{n}\frac{e^{-\beta_c E_n^0}}{Z^0(\beta_c)}e^{iu (E_n^1-E_n^0)}\\
&=\sum_{n}\frac{e^{-[\beta_c-iu(\epsilon^{-1}-1) ]E_n^0}}{Z^0(\beta_c)}\\
&=\frac{Z^0[\beta_c-iu(\epsilon^{-1}-1)]}{Z^0(\beta_c)},
 \end{split}
\end{gather}
and
\begin{gather}
 \begin{split}
\chi_h(u)&=\sum_{n}\frac{e^{-\beta_h E_n^1}}{Z^1(\beta_h)}e^{iu (E_n^0-E_n^1)}\\
&=\sum_{n}\frac{e^{-[\beta_c-iu(\epsilon-1) ]E_n^1}}{Z^1(\beta_h)}\\
&=\frac{Z^1[\beta_h-iu(\epsilon-1)]}{Z^1(\beta_h)}.
 \end{split}
\end{gather}
Then, it follows from Eq.~(\ref{b1}) that the joint characteristic function $\chi(u,v)$ reads
\begin{gather}
 \begin{split}
&\chi(u,v)=\\
&\frac{Z^0[\beta_c+iv-i(u-v)(\epsilon^{-1}-1)]Z^1[\beta_h-iv-iu(\epsilon-1)]}{Z^0(\beta_c)Z^1(\beta_h)}.
 \end{split}
\end{gather}

 \renewcommand{\theequation}{C\arabic{equation}}

 \setcounter{equation}{0}

 \section*{Appendix C: The joint characteristic function for a quantum harmonic oscillator heat engine}

For a harmonic oscillator with time-dependent frequency in an adiabatic
 process during time $[0,\tau]$, the Hamiltonian is
\begin{equation}
H(t)=\frac{p^2}{2m}+\frac{1}{2}m\omega(t)^2x^2.
\end{equation}
Then, the characteristic functions of work $\chi_{c}(u),\chi_{h}(u)$ (see Appendix B) reads~\cite{no2008, zy2019}
\begin{widetext}
\begin{equation}
\chi_{c}(u)=2\sinh\left(\frac{\beta\omega_{0}}{2}\right)\left\{ 2\cos(u\omega_{1})\cos[(u-i\beta_{c})\omega_{0}]+2Q_{c}\sin(u\omega_{1})\sin[(u-i\beta_{c})\omega_{0}]-2\right\} ^{-\frac{1}{2}},\label{eq:12-1}
\end{equation}
\begin{equation}
\chi_{h}(u)=2\sinh\left(\frac{\beta\omega_{1}}{2}\right)\left\{ 2\cos(u\omega_{0})\cos[(u-i\beta_{h})\omega_{1}]+2Q_{h}\sin(u\omega_{0})\sin[(u-i\beta_{h})\omega_{1}]-2\right\} ^{-\frac{1}{2}},\label{eq:13}
\end{equation}
\end{widetext}
where
\begin{equation}
Q_{c}=\frac{\omega_{1}}{2\omega_{0}}\left[y_{1}(\tau_c)^{2}+y_{2}(\tau_c)^{2}+\frac{\dot{y}_{1}(\tau_c)^{2}+\dot{y}_{2}(\tau_c)^{2}}{\omega_{1}^{2}}\right],\label{eq:Qdefinition}
\end{equation}
the overhead dot denotes the time derivative, %$[0,\tau]$ denotes the driving time of the adiabatic process, $\omega_{i(f)}$ denotes the frequency of the oscillator at time $t=0(\tau)$,
$y_{1}$ and $y_{2}$ are
the two general solutions of the classical harmonic
oscillator, i.e.,
\begin{equation}
\ddot{y}(t)+\omega(t)^{2}y(t)=0,\label{eq:15}
\end{equation}
with the initial value $\left\{ y_{1}(0),y_{2}(0),\dot{y}_{1}(0),\dot{y}_{2}(0)\right\} =\left\{ 1,0,0,\omega_0\right\}$. Similarly, the expression of $Q_h$ is given by the replacement: $\omega_0\leftrightarrow\omega_1$, $t\in[0,\tau_c]\rightarrow t\in[\tau_c+t_h,\tau_c+t_h+\tau_h]$.
%Moreover, it is easy to check that $Q=\left\langle H(\tau)\right\rangle /\left\langle H\right\rangle _{adi}$\cite{jaramillo2019}, where $\left\langle H(\tau)\right\rangle$ denotes the internal energy of the oscillator at time $t=\tau$, and $\left\langle H\right\rangle _{adi}$ denotes the internal energy of the oscillator at the end of the quantum adiabatic driving.
%Similarly, for the nonadiabatic factor $Q_{h}$, the expression can
%be obtained from Eq. (\ref{eq:14}) straightforwardly.

Then, the expression of $\chi(u,v)$ is obtained by $\chi(u,v)=\chi_{c}(u,v)\chi_{h}(u,v)$,
where (Eq.~(\ref{b1}))
\begin{widetext}
\begin{align*}
\chi_{c}(u,v) & =\chi_{c}(u)|_{u\rightarrow u-v,\beta_{c}\rightarrow\beta_{c}+iv}\\
 & =2\sinh\left(\frac{\beta\omega_{0}}{2}\right)\left\{ 2\cos[(u-v)\omega_{1}]\cos[(u-i\beta_{c})\omega_{0}]+2Q_{c}\sin[(u-v)\omega_{1})\sin[(u-i\beta_{c})\omega_{0}]-2\right\} ^{-\frac{1}{2}},
\end{align*}
\begin{align*}
\chi_{h}(u,v) & =\chi_{h}(u)|_{\beta_{h}\rightarrow\beta_{h}-iv}\\
 & =2\sinh\left(\frac{\beta\omega_{1}}{2}\right)\left\{ 2\cos(u\omega_{0})\cos[(u-v-i\beta_{h})\omega_{1}]+2Q_{h}\sin(u\omega_{0})\sin[(u-v-i\beta_{h})\omega_{1}]-2\right\} ^{-\frac{1}{2}}.
\end{align*}
\end{widetext}

 \renewcommand{\theequation}{D\arabic{equation}}

 \setcounter{equation}{0}

 \section*{Appendix D: Explicit Time-dependence
of the non-adiabatic factor}

For the specific driving protocol in Eq.~(\ref{protocol}),
the time-dependence of the non-adiabatic factor in Eq.~(\ref{eq:Qtau}) can be directly calculated
from its definition (Eq.~(\ref{eq:Qdefinition}))~\cite{beau2016}. Here, we present another approach
with respect to the internal energy of the working substance. It follows from Ref.~\cite{lee2021} that the
evolution of the harmonic oscillator in an adiabatic process during time $t\in[0,\tau]$ can be described
by a linear differential equation as

\begin{equation}
\frac{d}{dt}\overrightarrow{\phi}(t)=\mathcal{M}(t)\overrightarrow{\phi}(t).\label{eq:evolution}
\end{equation}
Here,

\begin{equation}
\overrightarrow{\phi}(t)\equiv\left(\begin{array}{ccc}
\left\langle H(t)\right\rangle  & \left\langle L(t)\right\rangle  & \left\langle D(t)\right\rangle \end{array}\right)^{\mathrm{T}},
\end{equation}
$\langle\cdot\rangle$ denotes the ensemble average respect to the density matrix of the oscillator, $\mathrm{T}$ denotes the matrix transpose. $H(t)=p^{2}/(2m)+m\omega^{2}(t)x^{2}/2$, $L(t)=p^{2}/(2m)-m\omega^{2}(t)x^{2}/2$,
and $D(t)=\omega(t)(xp+px)/2$ are respectively the Hamiltonian, the
Lagrangian, and the generator of the scale transformation. The time-dependent matrix reads
\begin{equation}
\mathcal{M}(t)=\left(\begin{array}{cccc}
\dot{\omega}/\omega^{2} & -\dot{\omega}/\omega^{2} & 0 \\
-\dot{\omega}/\omega^{2} & \dot{\omega}/\omega^{2} & -2 \\
0 & 2 & \dot{\omega}/\omega^{2}
\end{array}\right).\label{eq:M}
\end{equation}
The general solution of Eq. (\ref{eq:evolution}) follows as

\begin{equation}
\overrightarrow{\phi}(\tau)=\mathscr{T}_{\leftarrow}\exp\left[\int_{0}^{\tau}\mathcal{M}(t)\mathrm dt\right]\overrightarrow{\phi}(0),\label{eq:evolutiontau}
\end{equation}
where $\mathscr{T}_{\leftarrow}$ denotes the time-ordered operation. For the specific protocol in Eq.~(\ref{protocol}),
%For further analysis, we specific the protocol to tune
%the frequency of the harmonic oscillator in the adiabatic process
%as {[}1911.03622, Chen 2019{]},
%
%\begin{equation}
%\omega=\omega(t)=\frac{\omega_{0}}{(\omega_{0}/\omega_{1}-1)\frac{t}{\tau}+1},
%\end{equation}
%which makes
the matrix $\mathcal{M}(t)$ is independent of $t$.
For the thermal equilibrium initial state, $\overrightarrow{\phi}(0)=\left(\begin{array}{cccc}
\left\langle H(0)\right\rangle  & 0 & 0 \end{array}\right)^{\mathrm{T}}$,  we find

\begin{equation}
\overrightarrow{\phi}(\tau)=\left(\begin{array}{c}
\frac{-\frac{\omega_{f}}{\omega_{i}}\left[\left(\frac{\omega_{f}}{\omega_{i}}\right)^{-\sqrt{1-a^{2}\tau^2}}+\left(\frac{\omega_{f}}{\omega_{i}}\right)^{\sqrt{1-a^{2}\tau^2}}-2a^{2}\tau^2\right]}{2\left(a^{2}\tau^2-1\right)}\\
\frac{\left(\frac{\omega_{f}}{\omega_{i}}\right)^{1-\sqrt{1-a^{2}\tau^2}}\left[1-\left(\frac{\omega_{f}}{\omega_{i}}\right)^{2\sqrt{1-a^{2}\tau^2}}\right]}{2\sqrt{1-a^{2}\tau^2}}\\
\frac{a\tau\left(\frac{\omega_{f}}{\omega_{i}}\right)^{1-\sqrt{1-a^{2}\tau^2}}\left[1-\left(\frac{\omega_{f}}{\omega_{i}}\right)^{\sqrt{1-a^{2}\tau^2}}\right]^{2}}{2\left(a^{2}\tau^2-1\right)}
\end{array}\right)\left\langle H(0)\right\rangle ,
\end{equation}
where $a=2\omega_{i}\omega_{f}/(\omega_{f}-\omega_{i})$. Thus, the internal energy of the system
at $t=\tau$ is

\begin{equation}
\left\langle H(\tau)\right\rangle =\frac{a^{2}\tau^2-\cos(\sqrt{a^{2}\tau^2-1}\ln(\omega_f/\omega_i))}{a^{2}\tau^2-1}\frac{\omega_{f}}{\omega_{i}}\left\langle H(0)\right\rangle
\end{equation}
Consequently, the non-adiabatic factor is obtained by~\cite{jaramillo2019}

\begin{equation}
Q(\tau)=\frac{\left\langle H(\tau)\right\rangle }{\left\langle H\right\rangle _{adi}}=1+\frac{\left(1-\cos(\sqrt{a^{2}\tau^2-1}\ln(\omega_f/\omega_i))\right)}{a^{2}\tau^2-1},\label{eq:Q(t)}
\end{equation}
where $\left\langle H\right\rangle _{adi}=\left\langle H(a\tau\rightarrow\infty)\right\rangle =\left\langle H(0)\right\rangle \omega_f/\omega_i$
is the internal energy of the system at the end of the process under quantum adiabatic
driving. In the short-time limit $a\tau\rightarrow0$, and long-time
limit $a\tau\rightarrow\infty$, it is easy to check that

\begin{equation}
\lim_{a\tau\rightarrow0}Q(\tau)=1+\frac{\left[\ln(\omega_f/\omega_i)\right]^{2}}{2},\lim_{a\tau\rightarrow\infty}Q(\tau)=1.
\end{equation}

\end{document}